\documentclass[10pt,twocolumn,twoside]{IEEEtran}
\usepackage{graphicx}
\usepackage{amsmath}
\usepackage{array}
\newcommand{\PreserveBackslash}[1]{\let\temp=\\#1\let\\=\temp}
\newcolumntype{C}[1]{>{\PreserveBackslash\centering}p{#1}}
\newcolumntype{R}[1]{>{\PreserveBackslash\raggedleft}p{#1}}
\newcolumntype{L}[1]{>{\PreserveBackslash\raggedright}p{#1}}
\usepackage[usenames]{color}
\usepackage{colortbl,booktabs}
\usepackage{tabularx}
\usepackage{multicol}
\usepackage{booktabs}
\usepackage[Symbol]{upgreek}
\usepackage{subfigure}
\usepackage{bm}
\usepackage{cite}
\usepackage{balance}
\usepackage[boxed]{algorithm2e}
\usepackage{threeparttable}
\usepackage{amsmath}
\usepackage{dcolumn}
\usepackage{multirow}
\usepackage{booktabs}
\usepackage{amsfonts}
\newcolumntype{d}[1]{D{.}{.}{#1}}

\begin{document}

\bibliographystyle{IEEEtran} 
\title{Low-Complexity Soft-Output Signal Detection Based on Gauss-Seidel Method for Uplink Multi-User Large-Scale MIMO Systems}

\author{Linglong Dai,~\IEEEmembership{Senior Member,~IEEE}, Xinyu Gao,~\IEEEmembership{Student Member,~IEEE}, \\ Xin Su,~\IEEEmembership{Member,~IEEE}, Shuangfeng Han,~\IEEEmembership{Member,~IEEE}, \\ Chih-Lin I,~\IEEEmembership{Senior Member,~IEEE}, and Zhaocheng Wang,~\IEEEmembership{Senior Member,~IEEE}

\thanks{Copyright (c) 2013 IEEE. Personal use of this material is permitted. However, permission to use this material for any other purposes must be obtained from the IEEE by sending a request to pubs-permissions@ieee.org.}
\thanks{L. Dai, X. Gao, X. Su, and Z. Wang are with Tsinghua National Laboratory
for Information Science and Technology (TNList), Department of Electronic Engineering, Beijing 100084, China (e-mail: daill@tsinghua.edu.cn).}
\thanks{S. Han and C. I are with Green Communication Research Center, China Mobile Research
Institute, Beijing 100053, China (e-mail: hanshuangfeng@chinamobile.com).}
\thanks{This work was supported by National Key Basic Research Program of
China (Grant No. 2013CB329203), National Natural Science Foundation
of China (Grant Nos. 61271266 and 61201185),  Science and Technology Foundation for Beijing Outstanding Doctoral Dissertation (Grant No. 2012T50093).}}

\maketitle
\begin{abstract}
For uplink large-scale MIMO systems, minimum mean square error (MMSE) algorithm is near-optimal but involves  matrix inversion with high complexity. In this paper, we propose to exploit the Gauss-Seidel (GS) method to iteratively realize the MMSE algorithm without the complicated matrix inversion. To further accelerate the convergence rate and reduce the complexity, we propose a diagonal-approximate initial solution to the GS method, which is much closer to the final solution than the traditional zero-vector initial solution. We also propose a approximated method to compute log-likelihood ratios (LLRs) for soft channel decoding with a negligible performance loss. The analysis shows that the proposed GS-based algorithm can reduce the computational complexity from ${{\cal O}({K^3})}$  to ${{\cal O}({K^2})}$, where ${K}$ is the number of users. Simulation results verify that the proposed algorithm outperforms the recently proposed Neumann series approximation algorithm, and achieves the near-optimal performance of the classical MMSE algorithm with a small number of iterations.
\end{abstract}

\vspace{+5mm}

\section{Introduction}\label{S1}

\IEEEPARstart Multiple-input multiple-output (MIMO) technology has been successfully applied to many communication systems, such as the 4th generation (4G) cellular system LTE-A~\cite{Dongwoon12}, IEEE 802.11n wireless LAN system~\cite{Skordoulis08}, etc. It is widely recognized as a promising key technology for future  wireless communications~\cite{federico14}. Unlike the traditional small-scale MIMO (e.g., at most 8 antennas in LTE-A), large-scale MIMO, which equips a very large number of antennas (e.g., 128 antennas or even more) at the base station (BS) to simultaneously serve multiple user equipments (UEs), is recently proposed~\cite{marzetta10}. It has been  theoretically proved that large-scale MIMO can achieve orders of increase in spectrum and energy efficiency~\cite{ngo11}.

However, realizing the attractive benefits of large-scale MIMO in practice faces some challenging problems, one of which is the practical signal detection algorithm in the uplink~\cite{rusek13} due to the increased multi-user interferences. The optimal detector is the maximum likelihood (ML) detector, but its complexity increases exponentially with the number of transmit antennas, which makes it impractical for large-scale MIMO systems. Some non-linear detection algorithms such as fixed-complexity sphere decoding (FSD)~\cite{barbero08} and tabu search (TS)~\cite{datta10} are proposed to achieve close-optimal performance with reduced complexity. However, their complexity is still unaffordable when the dimension of the MIMO systems is large or the modulation order is high~\cite{rusek13} (e.g., 128 antennas at the BS with 64 QAM modulation). To make a tradeoff between the performance and complexity, one can resort to low-complexity linear detection algorithms such as zero-forcing (ZF) and minimum mean square error (MMSE) with near-optimal performance for uplink multi-user large-scale MIMO systems~\cite{rusek13}, but these algorithms involve unfavorable matrix inversion with high complexity. Recently, the  Neumann series approximation algorithm was proposed to convert the matrix inversion into a series of matrix-vector multiplications~\cite{yin13} to reduce the complexity. However, only marginal reduction in complexity can be achieved.

In this paper, we propose a low-complexity near-optimal signal detection algorithm based on the Gauss-Seidel (GS) method~\cite{bjorck1996numerical} for large-scale MIMO systems. Firstly, based on the special property that the MMSE filtering matrix of large-scale MIMO systems is Hermitian positive definite, we propose a low-complexity signal detection algorithm, which utilizes GS method to iteratively realize the MMSE estimate without matrix inversion. Then, based on the fact that the MMSE filtering matrix is diagonally dominant for uplink large-scale MIMO systems,  we propose to use the diagonal component of the MMSE filtering matrix to obtain a diagonal-approximate initial solution to the GS method, which can accelerate the convergence rate. After that, we propose a approximated method to calculate the channel gain and the noise-plus-interference (NPI) variance for log-likelihood ratios (LLRs) computation, which also utilizes the diagonal dominant property of the MMSE filtering matrix.
We verify through simulation results that the proposed GS-based algorithm with the approximated method for LLRs computation can attain the near-optimal performance of the classical MMSE algorithm with a small number of iterations. To the best of our knowledge, this work is the first one to utilize the GS method for signal detection in uplink large-scale MIMO systems.

The rest of the paper is organized as follows. Section~\ref{S2} briefly introduces the system model. Section~\ref{S3} specifies the proposed low-complexity signal detection algorithm based on the GS method. The simulation results of the bit-error rate (BER) performance are shown in Section~\ref{S4}. Finally, conclusions are drawn in Section~\ref{S5}.

{\it Notation}: We use lower-case and upper-case boldface letters to denote vectors and matrices, respectively; ${( \cdot )^T}$, ${( \cdot )^H}$, ${( \cdot )^{ - 1}}$, and $\left|  \cdot  \right|$ denote the transpose, conjugate transpose, matrix inversion, and absolute operators, respectively; ${{\mathop{\rm Re}\nolimits} \{  \cdot \} }$ and ${{\mathop{\rm Im}\nolimits} \{  \cdot \} }$ denote the real part and imaginary part of a complex number, respectively; Finally, ${{\bf{I}}_N}$ represents the $ N \times N $ identity matrix.

\section{System Model}\label{S2}
We consider a uplink large-scale MIMO system employing ${N}$  antennas at the BS to simultaneously serve ${K}$ selected single-antenna UEs for communications, where we usually have ${N \gg K}$, e.g., ${N = 128}$ and ${K = 16}$ have been considered in~\cite{Hoydis}. The parallel transmitted bit streams from ${K}$ different users are first separately encoded by the channel encoder, and then mapped to constellation symbols by taking values from a energy-normalized modulation constellation ${{\cal Q}}$. Let ${{\bf{s}}}$ denote the ${K \times 1}$ transmitted signal vector containing the transmitted symbols from all ${K}$ users, and ${{\bf{H}} \in {\mathbb{C}^{N \times K}}}$  denote the flat Rayleigh fading channel matrix whose entries are  independent and identically distributed (i.i.d.) with zero mean and unit variance~\cite{ngo11}. Then the ${N \times 1}$  received signal vector ${{\bf{y}}}$ at the BS can be presented as
\begin{equation}\label{eq1}
{\bf{y}} = {\bf{Hs}} + {\bf{n}},
\end{equation}
where ${{\bf{n}}}$ is a ${N \times 1}$ additive white Gaussian noise (AWGN) vector whose entries follow ${{\cal CN}(0,\sigma ^2)}$.

The task of multi-user signal detection at the BS is to estimate the transmitted signal vector ${{\bf{s}}}$  from the received noisy signal vector ${{\bf{y}}}$ (note that the channel matrix ${\bf{H}}$ can be usually obtained through time-domain and/or frequency-domain training pilots~\cite{dai13, Gao14b}). The estimate of the transmitted signal vector ${{\bf{\hat s}}}$ achieved by the MMSE linear detection algorithm can be presented as
\begin{equation}\label{eq2}
{\bf{\hat s}} = {\left( {{{\bf{H}}^H}{\bf{H}} + {\sigma ^2}{{\bf{I}}_K}} \right)^{ - 1}}{{\bf{H}}^H}{\bf{y}} = {{\bf{W}}^{ - 1}}{\bf{\bar y}},
\end{equation}
where ${{\bf{\bar y}} = {{\bf{H}}^H}{\bf{y}}}$ can be interpreted as the matched-filter output of ${{\bf{y}}}$, and the MMSE filtering matrix ${{\bf{W}}}$ is denoted by
\begin{equation}\label{eq3}
{\bf{W}} = {\bf{G}} + {\sigma ^2}{{\bf{{I}}}_{K}},
\end{equation}
where ${{\bf{G}} = {{\bf{{H}}}^H}{\bf{{H}}}}$  presents the Gram matrix. After the estimation of the transmitted signal vector, the soft information LLRs can be extracted from the estimated results for soft-input channel decoding.  Let ${{\bf{E}} = {{\bf{W}}^{ - 1}}{\bf{G}}}$ denote the equivalent channel matrix and ${{\bf{U}} = {{\bf{W}}^{ - 1}}{{\bf{H}}^H}{({{\bf{W}}^{ - 1}}{{\bf{H}}^H})^H} = {{\bf{W}}^{ - 1}}{\bf{G}}{{\bf{W}}^{ - 1}}}$. Then, by combining (1) and (2), the MMSE estimate ${{\bf{\hat s}}}$ can be rewritten as
${{\bf{\hat s}} = {\bf{Es}} + {{\bf{W}}^{ - 1}}{{\bf{H}}^H}{\bf{n}} }$.
The estimate of the transmitted symbol for the  ${k}$th user (i.e., the ${k}$th element of ${{\bf{\hat s}}}$) can be presented as ${{\hat s_k} = {\mu _k}{s_k} + {\nu _k}}$, where ${s_k}$ denotes the ${k}$th element of the transmitted signal vector ${{\bf{s}}}$, ${{\mu _k} = {E_{kk}}}$ is the equivalent channel gain, and ${\nu _k^2 = \sum\limits_{m \ne k}^K {{{\left| {{E_{mk}}} \right|}^2}}  + {U_{kk}}{\sigma ^2}}$ denotes the NPI variance, ${{{E_{mk}}}}$ and ${{U_{mk}}}$ present the element of matrix ${{\bf{E}}}$ and ${\bf{U}}$ in the ${m}$th row and ${k}$th column, respectively. Then the max-log approximated LLR ${L_{k,b}}$ of bit ${b}$  for the ${k}$th user  can be obtained by~\cite{Wu14}
\begin{equation}\label{eq4}
{L_{k,b}}\! =\! {\gamma _k}\left( {\mathop {\min }\limits_{q \in S_b^0} {{\left| {\frac{{{{\hat s}_k}}}{{{\mu _k}}}\! -\! q} \right|}^2}\! -\! \mathop {\min }\limits_{q' \in S_b^1} {{\left| {\frac{{{{\hat s}_k}}}{{{\mu _k}}}\! -\! q'} \right|}^2}} \right),
\end{equation}
where ${{\gamma _k} = \mu_k^2/\nu_k^2}$ is the signal-to-interference-plus-noise ratio (SINR) for the ${k}$th user, ${S_b^0}$  and  ${S_b^1}$ are the sets containing the symbols from the modulation constellation  ${{\cal Q}}$, where the  ${b}$th bit of the symbol is 0 and 1, respectively.

It has been proved that MMSE linear detection algorithm is near-optimal for uplink multi-user large-scale MIMO systems~\cite{rusek13}. However, the MMSE algorithm inevitably involves complicated matrix inversion ${{{\bf{W}}^{{\bf{ - 1}}}}}$ to achieve the MMSE estimate, the channel gain, and the NPI variance, all of which are required to calculate the final LLRs for soft-input channel decoding. The computational complexity of matrix inversion is ${{\cal O}({K^3})}$, which is high since ${K}$ is usually large in uplink large-scale MIMO systems~\cite{Hoydis}.

\section{Low-complexity Soft-output Signal Detection For Uplink Large-scale MIMO}\label{S3}
In this section, We first propose a low-complexity signal detection algorithm which utilizes GS method to iteratively realize the MMSE estimate without matrix inversion. To further accelerate the convergence rate and reduce the complexity, we also propose a diagonal-approximate initial solution to the GS method. Then we propose a approximated method to compute the channel gain and the NPI variance for LLRs computation, which does not need to compute the exact matrix inversion. Finally, the complexity analysis of the proposed GS-based algorithm is provided to show its advantages over conventional algorithms.

\subsection{Signal detection algorithm based on Gauss-Seidel method}\label{S2.1}
For uplink large-scale MIMO systems, the channel matrix ${\bf{H}}$ is column full-rank and column asymptotically orthogonal~\cite{rusek13}, which guarantees that the MMSE filtering matrix ${\bf{W}}$ is Hermitian positive definite.
This special property inspires us to exploit the GS method~\cite{bjorck1996numerical} to iteratively solve (2) without matrix inversion. The GS method is used to solve the ${N}$-dimension linear equation ${{\bf{Ax}} = {\bf{b}}}$, where ${{\bf{A}}}$ is the ${N \times N}$ Hermitian positive definite matrix, ${{\bf{x}}}$ is the ${N \times 1}$ solution vector, and ${{\bf{b}}}$ is the ${N \times 1}$  measurement vector. Unlike the traditional method that directly computes  ${{{\bf{A}}^{ - 1}}{\bf{b}}}$ to obtain ${{\bf{x}}}$, the GS method can iteratively solve the equation ${{\bf{Ax}} = {\bf{b}}}$ with low complexity. Since the MMSE filtering matrix ${{\bf{W}}}$  is also Hermitian positive definite as mentioned above, we can decompose ${{\bf{W}}}$ as
\begin{equation}\label{eq9}
{\bf{W}} = {\bf{D}} + {\bf{L}} + {{\bf{L}}^H},
\end{equation}
where ${{\bf{D}}}$, ${{\bf{L}}}$, and ${{{\bf{L}}^H}}$ denote the diagonal component, the strictly lower triangular component, and the strictly upper triangular component of ${{\bf{W}}}$, respectively. Then we can exploit the GS method to estimate the transmitted signal vector ${{\bf{s}}}$ as below
\begin{equation}\label{eq10}
{{\bf{s}}^{(i)}} = {({\bf{D + L}})^{ - 1}}({\bf{\bar y}} - {{\bf{L}}^H}{{\bf{s}}^{(i-1)}}),\quad i = 1,2, \cdot  \cdot  \cdot
\end{equation}
where ${i}$ is the number of iterations, and ${{{\bf{s}}^{(0)}}}$ denotes the initial solution which will be discussed later in Section~\ref{S3}-B. Since ${({\bf{D + L}})}$ is a lower triangular matrix, we can obtain ${{{\bf{s}}^{(i)}}}$ with low complexity as will be addressed in Section~\ref{S3}-D. It is worth noting that the proposed GS-based algorithm is convergent for any initial solution since the MMSE filtering matrix ${{\bf{W}}}$ is Hermitian positive definite~\cite[Theorem 7.2.2]{bjorck1996numerical}.

\subsection{Diagonal-approximate initial solution}\label{S2.3}
Traditionally, due to no priori information of the final solution is available, the initial solution ${{{\bf{s}}^{(0)}}}$ in (6) is set as a zero-vector~\cite{bjorck1996numerical}, which is simple but usually far away  from the final solution. Although the initial solution doesn't influence the convergence, it plays an important role in the convergence rate and affects both computational complexity and detection accuracy when the number of iterations is limited. In this subsection, we propose a diagonal-approximate initial solution to the GS-based algorithm to achieve a faster convergence rate.

For uplink large-scale MIMO systems, the channel matrix ${\bf{H}}$ is asymptotically orthogonal when ${N \gg K}$~\cite{rusek13}, so we have
\begin{equation}\label{eq10}
\frac{{\bf{h}}_m^H{{\bf{h}}_k}}{N} \to 0,\quad m \ne k,\quad m,k = 1,2, \cdot  \cdot  \cdot , K,
\end{equation}
where ${{{\bf{h}}_m}}$ denotes the ${m}$th column vector of the channel matrix ${\bf{H}}$. This indicates that the MMSE filtering matrix ${{\bf{W}} = {{\bf{H}}^H}{\bf{H}} + {\sigma ^2}{{\bf{I}}_{K}}}$ is diagonally dominant for uplink large-scale MIMO systems.
Based on this principle, we can conclude that the matrix ${{{\bf{W}}^{ - 1}}}$ is also diagonally dominant.
Fig. 1 shows the normalized entries of the matrix ${{{\bf{W}}^{ - 1}}}$ for different values of ${N}$ when ${K}$ is fixed to 16, where ${W_{mk}^{ - 1}}$ and ${W_{\max }^{ - 1}}$ denote the ${m}$th row and ${k}$th column entry and the maximum entry of ${{{\bf{W}}^{ - 1}}}$, respectively. We can observe that the domination of the diagonal elements of ${{{\bf{W}}^{ - 1}}}$ becomes more obvious with the increasing value of ${N/K}$, and the difference between the diagonal matrix ${{{\bf{D}}^{ - 1}}}$ and the non-diagonal matrix ${{{\bf{W}}^{ - 1}}}$ becomes smaller. This special property inspires us to utilize ${{{\bf{D}}^{ - 1}}}$ to approximate ${{{\bf{W}}^{ - 1}}}$ with small error~\cite{yin13,Wu14,Wu13}. Then, the initial solution ${{{\bf{s}}^{(0)}}}$ in (6) can be approximately selected as
\begin{equation}\label{eq24}
{{\bf{s}}^{(0)}} = {{\bf{D}}^{ - 1}}{\bf{\hat y}}.
\end{equation}

\begin{figure}[h]
\begin{center}
\includegraphics[width=0.98\linewidth]{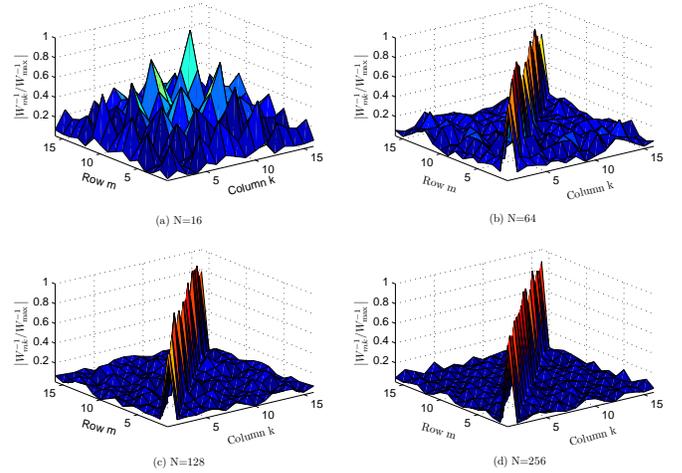}
\end{center}
\vspace{-3mm}
\caption{Normalized entries of ${{{\bf{W}}^{ - 1}}}$ for different values of ${N}$ when ${K}$ is fixed to 16.} \label{FIG1}
\end{figure}

Since the approximation error of ${{{\bf{s}}^{(0)}}}$ in (8) should be small as shown in Fig. 1, it is expected that the proposed diagonal-approximate initial solution will be closer to the final MMSE estimate ${{\bf{\hat s}}}$ compared to the traditional zero-vector initial solution. Therefore, a faster convergence rate can be achieved. Besides, note that the computational complexity to compute ${{{\bf{D}}^{ - 1}}}$ (or equivalently ${{{\bf{s}}^{(0)}}}$) is very low, since  ${{\bf{D}}}$ is a diagonal matrix.

\vspace*{0mm}
\subsection{Approximated method to compute LLRs}\label{S2.4}
\emph{1) Exact method}: Although the GS-based algorithm is originally designed to obtain the MMSE estimate ${{\bf{\hat s}}}$, it can be also utilized to obtain the estimate of the matrix inversion ${{{\bf{W}}^{ - 1}}}$. Combining (2) and (6), we set ${{\bf{\bar y}} = {{\bf{e}}_m}}$ where ${{{\bf{e}}_m}}$ denotes the ${m}$th ${K \times 1}$ unit vector, then the result of the GS-based algorithm ${\left( {{{{\bf{\hat w}}}_{{\rm{inv}}}}} \right)_m^{(i)}\! =\! {({\bf{D}}\! +\! {\bf{L}})^{\! -\! 1}}\left( {{{\bf{e}}_m}\! -\! {{\bf{L}}^H}\left( {{{{\bf{\hat w}}}_{{\rm{inv}}}}} \right)_m^{(i\! -\! 1)}} \right)}$ for ${i = 1,2, \cdot  \cdot  \cdot }$ will be the ${m}$th column of the estimate of the matrix ${{{\bf{W}}^{ - 1}}}$ in the ${i}$th iteration, where ${{\left( {{{{\bf{\hat w}}}_{\rm{inv}}}} \right)_m^{(0)}}}$ can be selected as the diagonal-approximate initial solution addressed in Section~\ref{S3}-B, i.e., ${\left( {{{{\bf{\hat w}}}_{{\rm{inv}}}}} \right)_m^{(0)} = {{\bf{D}}^{ - 1}}{{\bf{e}}_m}}$. Thus, the estimate ${{\left( {{{{\bf{\hat W}}}_{{\rm{inv}}}}} \right)^{(i)}}}$ of the matrix ${{{\bf{W}}^{ - 1}}}$ in the ${i}$th iteration can be achieved by
\begin{equation}\label{eq24}
\begin{array}{l}
{\left( {{{{\bf{\hat W}}}_{{\rm{inv}}}}} \right)^{(i)}} = {({\bf{D}} + {\bf{L}})^{ - 1}}\left( {{{\bf{I}}_K} - {{\bf{L}}^H}{{\left( {{{{\bf{\hat W}}}_{{\rm{inv}}}}} \right)}^{(i - 1)}}} \right),\quad \\
\quad \quad \quad \quad \quad \quad  \quad \quad i = 1,2, \cdot  \cdot  \cdot
\end{array}
\end{equation}
Then, by replacing the matrix ${{{\bf{W}}^{ - 1}}}$ by the estimated matrix ${{\left( {{{{\bf{\hat W}}}_{\rm{inv}}}} \right)^{(i)}}}$, we have ${{{\bf{\hat E}}^{(i)}} = {\left( {{{{\bf{\hat W}}}_{\rm{inv}}}} \right)^{(i)}}{\bf{G}}}$ and ${{{\bf{\hat U}}^{(i)}} = {\left( {{{{\bf{\hat W}}}_{\rm{inv}}}} \right)^{(i)}}{\bf{G}}{\left( {{{{\bf{\hat W}}}_{\rm{inv}}}} \right)^{(i)}}}$, and the equivalent channel gain ${\hat \mu _k^{(i)}}$ and NPI variance ${{\left( {\hat \nu _k^{(i)}} \right)^2}}$ achieved by the GS-based algorithm in the ${i}$th iteration can be presented as
\begin{equation}\label{eq24}
\hat \mu _k^{(i)} = \hat E_{kk}^{(i)},
\end{equation}
\begin{equation}\label{eq24}
{\left( {\hat \nu _k^{(i)}} \right)^2} = \sum\limits_{m \ne k}^K {{{\left| {\hat E_{mk}^{(i)}} \right|}^2}}  + \hat U_{kk}^{(i)}{\sigma ^2},
\end{equation}
Substituting (6), (10), and (11) into (4), we can obtain the exact mag-log LLRs for soft-input channel decoding.

\emph{2) Approximated method}: The exact method above  can compute the exact max-log LLRs to produce a good BER performance, but it inevitably involves  the calculation of ${{{{\left( {{{{\bf{\hat W}}}_{\rm{inv}}}} \right)}^{(i)}}}}$, which requires ${K}$ times of the GS method with the complexity ${{\cal O}({K^2})}$ for each time. Therefore, although the proposed GS-based algorithm can obtain the MMSE estimate ${{\bf{\hat s}}}$ with low complexity ${{\cal O}({K^2})}$, the exact method to compute LLRs still suffers from the complexity as high as ${{\cal O}({K^3})}$. To solve this problem, we propose
an approximated method inspired by~\cite{Wu13} to calculate the channel gain and NPI variance for LLRs computation, which can avoid the complicated matrix inversion.

Since ${{{\bf{W}}^{ - 1}}}$ is diagonal dominant for uplink large-scale MIMO systems as we have verified in Section~\ref{S3}-B, we can utilize the diagonal matrix ${{{\bf{D}}^{ - 1}}}$ to approximate ${{{\bf{W}}^{ - 1}}}$ with small error~\cite{yin13,Wu14,Wu13}. Then the approximated channel gain ${\tilde \mu _k}$ and the approximated NPI variance ${\tilde \nu _k^2}$ can be achieved by
\begin{equation}\label{eq24}
{\tilde \mu _k} = {\tilde E_{kk}},
\end{equation}
\begin{equation}\label{eq24}
\tilde \nu _k^2 = \sum\limits_{m \ne k}^K {{{\left| {{{\tilde E}_{mk}}} \right|}^2}}  + {\tilde U_{kk}}{\sigma ^2},
\end{equation}
where ${{\bf{\tilde E}} = {{\bf{D}}^{ - 1}}{\bf{G}}}$ and ${{\bf{\tilde U}} = {{\bf{D}}^{ - 1}}{\bf{G}}{{\bf{D}}^{ - 1}}}$. Substituting (6), (12), and (13) in (4), we can obtain the approximated  max-log  LLRs. Since ${{{\bf{D}}^{ - 1}}}$ is a diagonal matrix, the computation of ${{\bf{\tilde E}}}$ and ${{\bf{\tilde U}}}$ involves low complexity. Besides, since the MMSE estimate ${{\bf{\hat s}}}$ can be obtained without matrix inversion, the overall computational complexity to compute LLRs can be significantly reduced as will be quantified in the following subsection.

It is worth pointing out that the method proposed in~\cite{Wu13} also utilizes ${{{\bf{D}}^{ - 1}}}$ to approximate ${{{\bf{W}}^{ - 1}}}$, but it simplifies the LLRs computation by first computing a conjugate gradient matrix with low complexity, which is then used to compute LLRs, while our method directly utilizes ${{{\bf{D}}^{ - 1}}}$ to obtain LLRs.

\subsection{Computational complexity analysis }\label{S2.4}
Since both the MMSE algorithm and the proposed GS-based algorithm need to compute the Gram matrix ${{\bf{G}} = {{\bf{H}}^H}{\bf{H}}}$ (or equivalently ${{\bf{W}} = {\bf{G}} + {\sigma ^2}{{\bf{{I}}}_{K}}}$) and the matched-filter output ${{\bf{\bar y}}}$, we focus on the complexity of the LLRs computation, and evaluate it in terms of the required number of (complex) multiplications~\cite{gao2014low}.
It can be found from (4) that the computational complexity of the proposed GS-based algorithm to obtain LLRs comes from three parts:

1) The first one comes from the diagonal-approximate initial solution (8) addressed in Section III-B, which involves the computation of ${{{\bf{D}}^{ - 1}}}$ and a multiplication of the ${K \times K}$ diagonal matrix ${{{\bf{D}}^{ - 1}}}$ and the ${K \times 1}$ vector ${{\bf{\bar y}}}$. Therefore the required number of complex multiplications is ${2K}$.

2) The second one originates from solving the linear equation (6). Considering the definition of ${{\bf{D}}}$ and ${{\bf{L}}}$ in (5), the solution can be presented as
\begin{equation}\label{eq25}
\begin{array}{l}
s_m^{(i)} = \frac{1}{{{W_{mm}}}}({{\bar y}_m} - \sum\limits_{k < m} {{W_{mk}}s_k^{(i)} - \sum\limits_{k > m} {{W_{mk}}s_k^{(i-1)}} } ), \\
\quad \quad \quad \quad \quad \quad m,k = 1,2, \cdot  \cdot  \cdot, K,
\end{array}
\end{equation}
where ${s_m^{(i)}}$, ${s_m^{(i-1)}}$, and ${{\bar y_m}}$  denote the  ${m}$th element of  ${{{\bf{s}}^{(i)}}}$, ${{{\bf{s}}^{(i-1)}}}$, and ${{\bf{\bar y}}}$, respectively, and ${{W_{mk}}}$ denotes the element of ${{\bf{W}}}$ in the ${m}$th row and  ${k}$th column. It is clear that the required number of complex multiplications to compute ${s_m^{(i)}}$ is ${K}$. Since there are ${K}$ elements in vector ${{{\bf{s}}^{(i)}}}$, solving the equation (6) requires ${{iK^2}}$ times of complex multiplications.

3) The third one is from the computation of channel gain and NPI variance. It can be found from (12) and (13) that for the proposed approximated method  to compute LLRs, it requires to calculate two parts, i.e., all the elements of the matrix ${{\bf{\tilde E}}}$ and the diagonal elements of the matrix ${{\bf{\tilde U}}}$. Due to ${{\bf{\tilde E}} = {{\bf{D}}^{ - 1}}{\bf{G}}}$, and the diagonal matrix ${{{\bf{D}}^{ - 1}}}$ has been obtained when we use the diagonal-approximate initial solution, the required number of complex multiplications of the first part is ${{K^2}}$. For the second part, we only need the diagonal elements of the matrix ${{\bf{\tilde U}}}$, which can be presented as ${{\tilde U_{kk}} = D_{kk}^{ - 2}{G_{kk}}}$ for ${k = 1,2, \cdot  \cdot  \cdot, K}$, where ${D_{kk}^{ - 1}}$ and ${{G_{kk}}}$ denote the ${k}$th diagonal element of ${{{\bf{D}}^{ - 1}}}$ and ${{\bf{G}}}$, respectively. Thus, the required number of complex multiplications of the second part is as small as ${2K}$.

To sum up, the overall required number of complex multiplications by the proposed GS-based algorithm is ${(i + 1){K^2} + 4K}$, so the computational complexity is ${{\cal O}({K^2})}$ for arbitrary number of iterations.

\begin{figure}[tp]
\begin{center}
\includegraphics[width=1\linewidth]{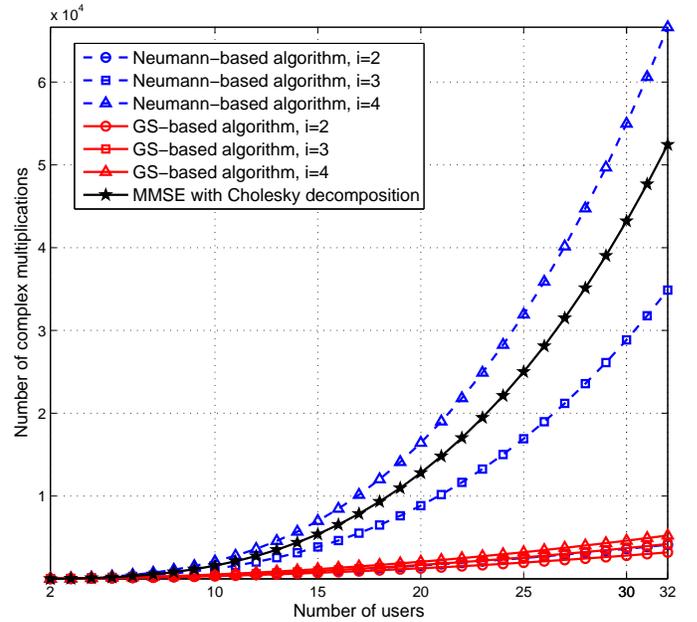}
\end{center}
\vspace{-3mm}
\caption{Complexity comparison against number of users.} \label{FIG2}
\end{figure}

Fig. 2 compares the complexity of the Neumann-based algorithm~\cite{Wu14} and the proposed GS-based algorithm, whereby the MMSE algorithm with Cholesky decomposition is also included as a baseline for comparison~\cite{Wu14}. Note that all these three algorithms utilize the approximated method to compute the LLRs as described in Section~\ref{S3}-C. It shows that the Neumann-based algorithm has lower complexity than the MMSE algorithm with Cholesky decomposition when ${i \le\ 3}$, especially when ${i=2}$ with the complexity ${{\cal O}({K^2})}$. However, when ${i \ge 4}$, the complexity of the Neumann-based algorithm is ${{\cal O}\left( {(i - 2){K^3}} \right)}$~\cite{Wu14}, which is even higher than that of  the MMSE algorithm. Since usually large value of ${i}$  is required to ensure the final approximation performance as will be verified later by simulation results in Section~\ref{S4}, the reduction in complexity achieved by the Neumann-based algorithm is marginal. By contrast, since ${K}$ is usually large for large-scale MIMO systems (e.g., ${K=16}$ in~\cite{rusek13}), we can observe that the proposed GS-based algorithm can evidently reduce the complexity from ${{\cal O}({K^3})}$  to ${{\cal O}({K^2})}$ for arbitrary number of iterations. Even for ${i=2}$, the proposed algorithm enjoys a lower complexity than the Neumann-based algorithm. Moreover, as shown in Fig. 2, the proposed GS-based algorithm will lead to more significant reduction in complexity when the dimension of MIMO system becomes larger, which means that the proposed algorithm with low complexity is quite suitable for  large-scale MIMO systems.

Additionally, we can observe from (14) that the computation of ${s_m^{(i)}}$  utilizes ${s_k^{(i)}}$ for ${k = 1,2, \cdot  \cdot  \cdot ,m - 1}$ in the current ${i}$th iteration and ${s_l^{(i-1)}}$ for ${l = m + 1,m+2, \cdot  \cdot  \cdot ,K}$ in the previous ${(i-1)}$th iteration. This characteristic of the GS method will lead to the GS-based algorithm hard to be parallelized, however it can bring two other benefits. Firstly, after ${s_m^{(i)}}$  has been obtained, we can use it to overwrite ${s_m^{(i-1)}}$ which is useless in the next computation of  ${s_{m + 1}^{(i)}}$. Consequently, only one storage vector of size ${K \times 1}$ is required; Secondly, when ${i}$ increases, the solution to (6) becomes closer to the final MMSE estimate  ${{\bf{\hat s}}}$. Thus, ${s_m^{(i)}}$ can exploits the elements of ${s_k^{(i)}}$ for ${k = 1,2, \cdot  \cdot  \cdot ,m - 1}$  that have already been  computed in the current iteration to produce more reliable result than the conventional algorithm, which only utilizes all the elements of ${{{\bf{s}}^{(i-1)}}}$ in the previous iteration. Thus, a faster convergence rate can be expected, and the required number of iterations to achieve a certain estimate accuracy becomes smaller. Based on these facts, the overall complexity of the proposed algorithm can be reduced further.

\section{Simulation Results}\label{S4}
The simulation results of BER performance against the signal-to-noise ratio (SNR) are provided to compare the GS-based algorithm with the recently proposed Neumann-based algorithm~\cite{Wu14}. The BER performance of the classical MMSE algorithm with Cholesky decomposition is also included as the benchmark for comparison. In all simulations, we consider the modulation scheme of 64 QAM, and the rate-1/2 industry standard convolutional code with ${[{133_o}\;{171_o}]}$ polynomial. At the receiver, LLRs are extracted from the detected signal for soft-input Viterbi decoding.  Note that the SNR is defined at the receiver~\cite{Wu14}.

\begin{figure}[tp]
\begin{center}
\includegraphics[width=0.98\linewidth]{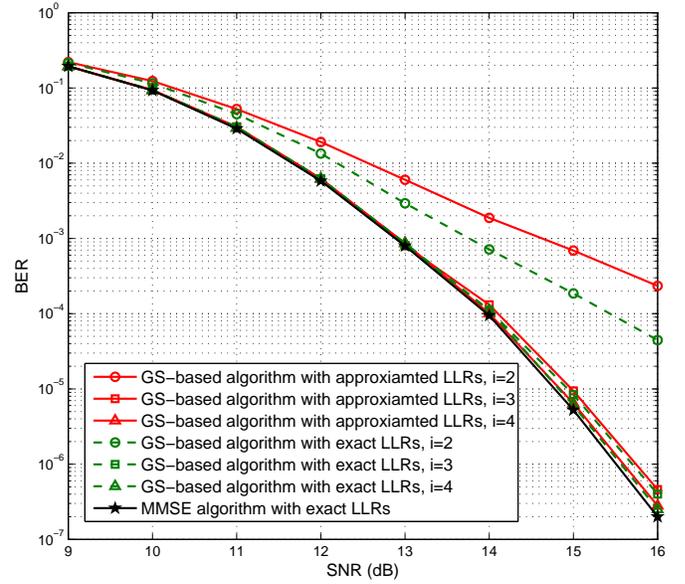}
\end{center}
\vspace{-3mm}
\caption{BER performance comparison between the exact method and the proposed approximated method to compute LLRs.} \label{FIG3}
\end{figure}

\begin{figure}[tp]
\begin{center}
\includegraphics[width=0.98\linewidth]{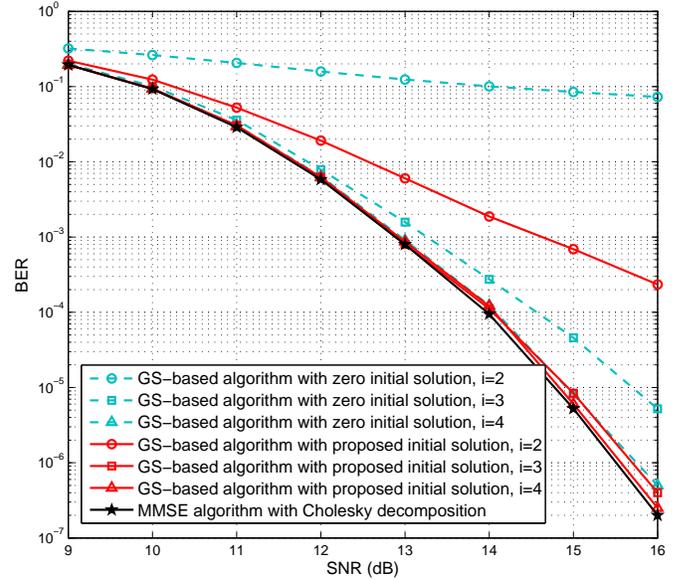}
\end{center}
\vspace{-3mm}
\caption{BER performance comparison between the conventional zero initial solution and the proposed initial solution.} \label{FIG4}
\end{figure}

Firstly, we consider the uncorrelated Rayleigh fading channel. Fig. 3 shows the BER performance comparison between the GS-based algorithm with the exact method and the approximated method to compute LLRs, when ${N \times K = 128 \times 16}$. Note that ${i}$ denotes the number of iterations and we choose the diagonal-approximate initial solution. We can observe that compared to the exact method to compute LLRs involving high complexity, the proposed approximated method can achieve a satisfying performance when the number of iterations ${i}$ is relatively large (e.g., ${i \ge 3}$). For example, when ${i=3}$, the difference between the exact method and the approximated method is within 0.1 dB.

Fig. 4 compares the BER performance between the GS-based algorithm with the conventional zero-vector initial solution and the proposed diagonal-approximate initial solution, when ${N \times K = 128 \times 16}$ and the approximated method to compute LLRs is employed. It is clear that the proposed diagonal-approximate initial solution can accelerate the convergence rate. When ${i = 3}$, the GS-based algorithm with diagonal-approximate initial solution has almost the same performance as that with the conventional zero-vector initial solution when ${i=4}$, which means the overall complexity of the proposed algorithm can be reduced further.

\begin{figure}[tp]
\begin{center}
\includegraphics[width=0.98\linewidth]{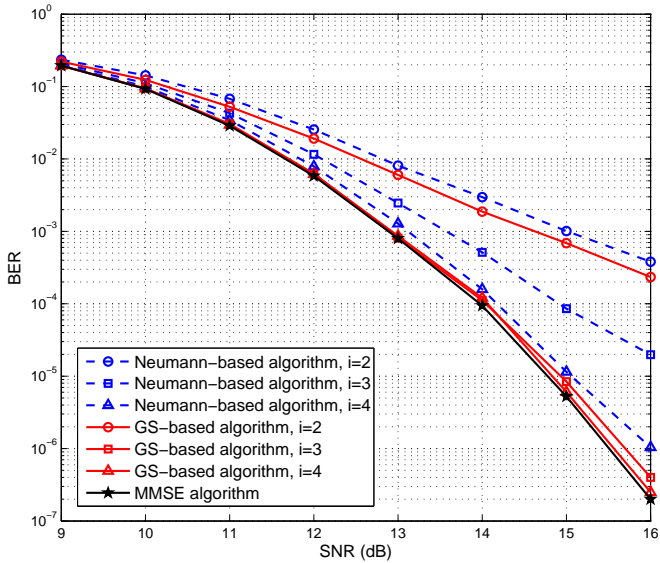}
\end{center}
\vspace{-3mm}
\caption{BER performance comparison between the conventional Neumann-based algorithm and the proposed GS-based algorithm.} \label{FIG5}
\end{figure}

\begin{figure}[tp]
\begin{center}
\includegraphics[width=0.98\linewidth]{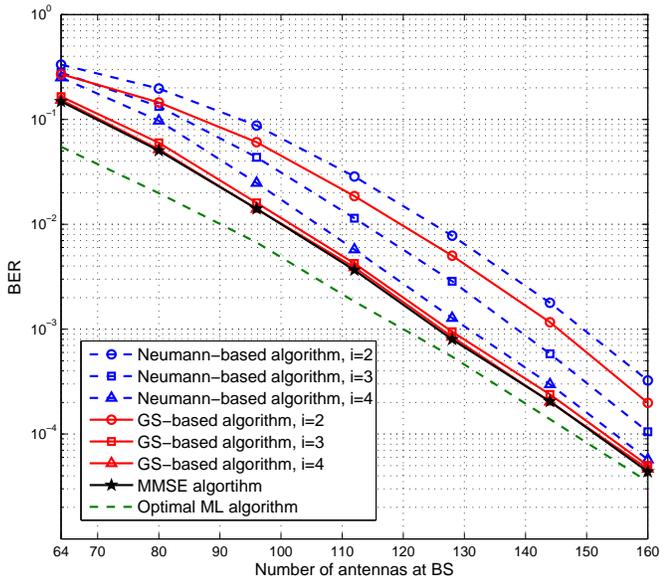}
\end{center}
\vspace{-3mm}
\caption{BER performance comparison against the number of antennas at BS.} \label{FIG6}
\end{figure}

Fig. 5 shows the BER performance comparison between the conventional  Neumann-based algorithm~\cite{Wu14} and the proposed GS-based algorithm, when ${N \times K = 128 \times 16}$. Note that we choose the diagonal-approximate initial solution and the proposed approximated method to compute LLRs for the GS-based algorithm. It is clear that with the increased number of iterations, the BER performance of both algorithms becomes closer to the MMSE algorithm. However, the GS-based algorithm outperforms the conventional one when the same number of iterations is used. As we can observe from Fig. 5, when ${i=3}$, the SNR required by the GS-based algorithm to achieve the BER of ${{10^{ - 4}}}$ is 14 dB, while for the Neumann-based algorithm, the required SNR is 15 dB.

In addition, in Fig. 6 we also provide the simulation results about the BER performance of the proposed GS-based algorithm against the number of antennas at BS (${N}$) when a fixed number of users ${K=16}$ is considered. Note that SNR = 13 dB is adopted.
We can observe that the performance of the MMSE algorithm improves when ${N}$ increases, and the GS-based algorithm can achieve the exact performance of the MMSE algorithm with a small number of iterations (i.e., ${i=4}$) regardless of the value of ${N}$. By contrast, although the performance of the Neumann-based algorithm also improves with the increasing value of ${N}$, it still suffers a non-negligible performance loss, which further verifies that the proposed  GS-based algorithm outperforms the conventional Neumann-based algorithm in large-scale MIMO systems. More importantly, we can also observe from Fig. 6 that the proposed GS-based algorithm is near-optimal compared to the optimal ML algorithm, since when ${N \gg K}$ (e.g., ${N/K=8}$) the performance of the GS-based algorithm with ${i=4}$ is close to that of the optimal ML algorithm.

\begin{figure}[tp]
\begin{center}
\includegraphics[width=0.95\linewidth]{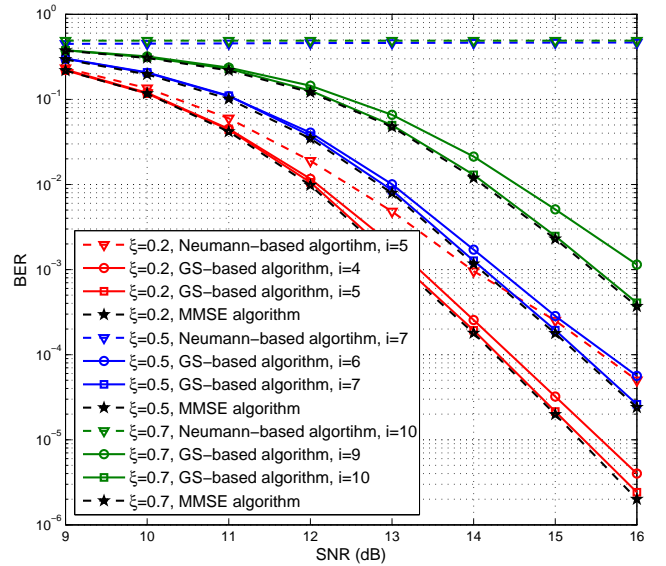}
\end{center}
\vspace{-3mm}
\caption{BER performance comparison for different values of correlated magnitude.} \label{FIG7}
\end{figure}

Finally, as the spatial correlation of MIMO channels plays a crucial role in the performance of realistic MIMO systems, we show in Fig. 7 how the channel correlation affects the performance of the proposed GS-based algorithm. Note that we adopt the exponential correlation model described in~\cite{Godana13}, and ${\xi }$ (${0 \le \xi  \le 1}$) denotes the correlated factor between two adjacent antennas.
We can observe that the performance of the classical MMSE algorithm degrades when the channel correlation becomes serious, which is consistent with the theoretical analysis in~\cite{Godana13}, and the GS-based algorithm can still converge to the MMSE algorithm without obvious performance loss. However, the required number of iterations by the  proposed GS-based algorithm to converge becomes larger with an increasing value of ${\xi }$ (e.g., ${i = 7}$ when ${\xi=0.5}$, but ${i=10}$ when ${\xi=0.7}$), which means more serious channel correlation will lead to slower convergence rate. However, the GS-based algorithm can still enjoy a lower complexity than the Neumann-based algorithm and the MMSE algorithm with Cholesky decomposition.

\section{Conclusions}\label{S5}
In this paper, by fully exploiting the special characteristics of uplink large-scale MIMO systems, we propose a low-complexity near-optimal signal detection algorithm based on the GS method. To reduce the complexity further, we propose a diagonal-approximate initial solution to the GS-based algorithm which is close to the final solution to accelerate the convergence rate. We also propose a approximated method to compute LLRs with low complexity for soft-input channel decoding. Analysis shows that the proposed algorithm can reduce the complexity from ${{\cal O}({K^3})}$ to ${{\cal O}({K^2})}$. It is verified that the proposed algorithm outperforms the conventional method, and achieves the near-optimal performance of the classical MMSE algorithm with a small number of iterations. Additionally, the idea of using the GS method to efficiently solve the complicated matrix inversion can be applied to other signal processing problems involving matrix inversion of large size in wireless communications, such as the downlink precoding in large-scale MIMO systems.


\bibliography{IEEEabrv,Gao1Ref}

\balance

\end{document}